Earthquake Swarm Activity in the Tokara Islands (2025): Statistical Analysis Indicates Low Probability of Major Seismic Event

Tomokazu Konishi

Faculty of Bioresource Sciences, Akita Prefectural University

ABSTRACT

The Tokara Islands, a volcanic archipelago situated south of Japan's main islands, has experienced sustained earthquake swarm activity in 2025. Public concern has emerged regarding potential triggering of the anticipated Nankai Trough earthquake, which the Japan Meteorological Agency has dismissed; however, the underlying mechanisms of this seismic activity remain inadequately explained. This study employs Exploratory Data Analysis (EDA) to characterise the statistical properties of the current earthquake swarm and compare them with historical patterns. The frequency and magnitude distributions of the 2025 swarm demonstrate remarkable similarity to two previous swarms that occurred in 2021. Both the current episode and the 2021 events coincided with volcanic activity at Suwanose Island, located approximately 10 km from the epicentral region, suggesting a causal relationship between magmatic processes and seismic activity. Statistical analysis reveals that the earthquake swarm exhibits exceptionally low magnitude scale ($\sigma = 0.37$) and moderate mean magnitude ($\mu = 2.8$), characteristics consistent with magma-driven seismicity rather than tectonic stress accumulation. These parameters contrast markedly with pre-seismic conditions observed before the 2011 Tohoku earthquake, where both σ and μ were substantially elevated (σ = 1.2, μ = 3.6). The exponential distribution of earthquake intervals (λ = 1/0.19 hours) further supports the interpretation of magma-induced seismicity. Our findings indicate that the current seismic activity represents localised volcanic-related processes rather than precursory behaviour associated with major tectonic earthquakes. The statistical signatures provide no evidence for imminent large-scale seismic events, including the Nankai Trough earthquake system. These results demonstrate the utility of statistical seismology in distinguishing between volcanic and tectonic seismic processes for hazard assessment purposes.

## INTRODUCTION

The application of Exploratory Data Analysis (EDA), a contemporary statistical methodology, has revealed distinct distributional characteristics for earthquake intervals and magnitudes (Konishi, 2025). Specifically, earthquake intervals demonstrate exponential distribution, whilst magnitudes follow normal distribution, thereby disproving the long-established Gutenberg-Richter law (Gutenberg and Richter, 1944). This discovery enables precise measurement of parameters characterising seismic activity within specific temporal and spatial contexts, facilitating comparative analysis and evaluation.

Recently, a month-long earthquake swarm has occurred in the Tokara Islands, a volcanic archipelago situated in the southern seas, considerably distant from Japan's main islands (JMA, 2025a, b). This has happened over 2000 times to date. This region lies near the Kikai Caldera, which experienced a supereruption approximately 7,300 years ago (JAMSTEC, 2025; Yamada et al, 2020), and represents the southern terminus of the Nankai Trough (JMA, 2025c), a zone of significant seismic concern. Consequently, speculation regarding catastrophic eruptions and earthquakes has proliferated across social media platforms, prompting official denials from the Japan Meteorological Agency (JMA). However, a comprehensive explanation of the current seismic activity remains absent. This study attempts to provide such an explanation through the application of EDA.

The region has historically been characterised by frequent seismic activity. Several islands host active volcanoes, particularly Suwanose Island, which lies in proximity to the epicentral area and exhibits frequent eruptive behaviour. The relationship between volcanic and seismic phenomena is well-established; the energy released during tectonic plate movement may contribute to magma generation  (JMA, 2025d; Survey, 2025). Seismic events may enhance volcanic eruption likelihood through induced strain (Nishimura, 2021), whilst conversely, eruptions may trigger seismic activity.

However, the relationship between eruptions and earthquakes at Suwanose Island demonstrates temporal complexity (JMA, 2025e). During 2021, earthquake swarms occurred in April and December, about 300 times each, whilst eruptive activity continued throughout this period. The April swarm commenced approximately one week following a significant eruption, whereas the December swarm occurred several months subsequently. The spatial distribution of seismic events during this period exhibited similarities to the current earthquake swarm pattern. Presently, Suwanose

Island has maintained volcanic activity since June 2023, with increased magma accumulation reported in May 2025.

METHODS

**Data Collection**

The most recent data on earthquake occurrence times and magnitudes were obtained from the website (JMA, 2025a). Epicentre locations were also acquired from the website (JMA, 2025f). Historical data were obtained from JMA public datasets (JMA, 2025b), which included epicentre information.

**Data Distribution Confirmation and Parameter Estimation**

All statistical analyses were conducted using R, a statistical computing environment (Konishi, 2025; R Core Team, 2025). Inter-earthquake intervals were calculated by determining the temporal difference between consecutive events in chronological order. Whilst no specific geographical region was designated for the overall analysis, the data presented in Figure 3 are restricted to the Tokara Islands area. The data were sorted and compared with equivalent numbers of theoretical distributions. Parameters were estimated from these linear relationships using the robust R line function.

RESULTS

The frequency of this earthquake swarm is exceptionally high, averaging approximately once every 0.19 hours (Figure 1A). The quantile-quantile (QQ) plot exhibits significant deviation from linearity due to this swarm activity, which occurs 36 times more frequently than the previous baseline frequency of once every 6.8 hours. Prior to the earthquake swarm, the QQ plot demonstrated linearity consistent with exponential distribution (Figure S1A, Appendix A). Such earthquake swarms are frequently observed following major seismic events or during periods of volcanic activity. For instance, similar patterns were documented following the 2000 Miyake Island eruption, where the QQ plot exhibited comparable deviation from linearity (Figure 1B). In that case, seismic frequency increased from once every 8.3 hours pre-eruption to once every 0.34 hours post-eruption.

The magnitudes recorded during this earthquake cluster were substantially lower than baseline values (Figure 1C). The graph is curved because there are two different phases (compare with Figure S1B). The magnitude distribution was similarly affected by this clustering, with the scale ($\sigma$) decreasing from $\sigma = 1.2$ to $\sigma = 0.37$. This contrasts markedly with the Miyake Island case, where $\sigma$ increased from an initial value of 0.65 to 1.9. This disparity likely reflects fundamental differences in the causal mechanisms underlying the Miyake Island and Tokara seismic events. At Miyake Island, a significant earthquake occurred almost simultaneously with the eruption (Figure 1D), whereas at Tokara, the temporal relationship between eruptive and seismic activities was not necessarily synchronous.

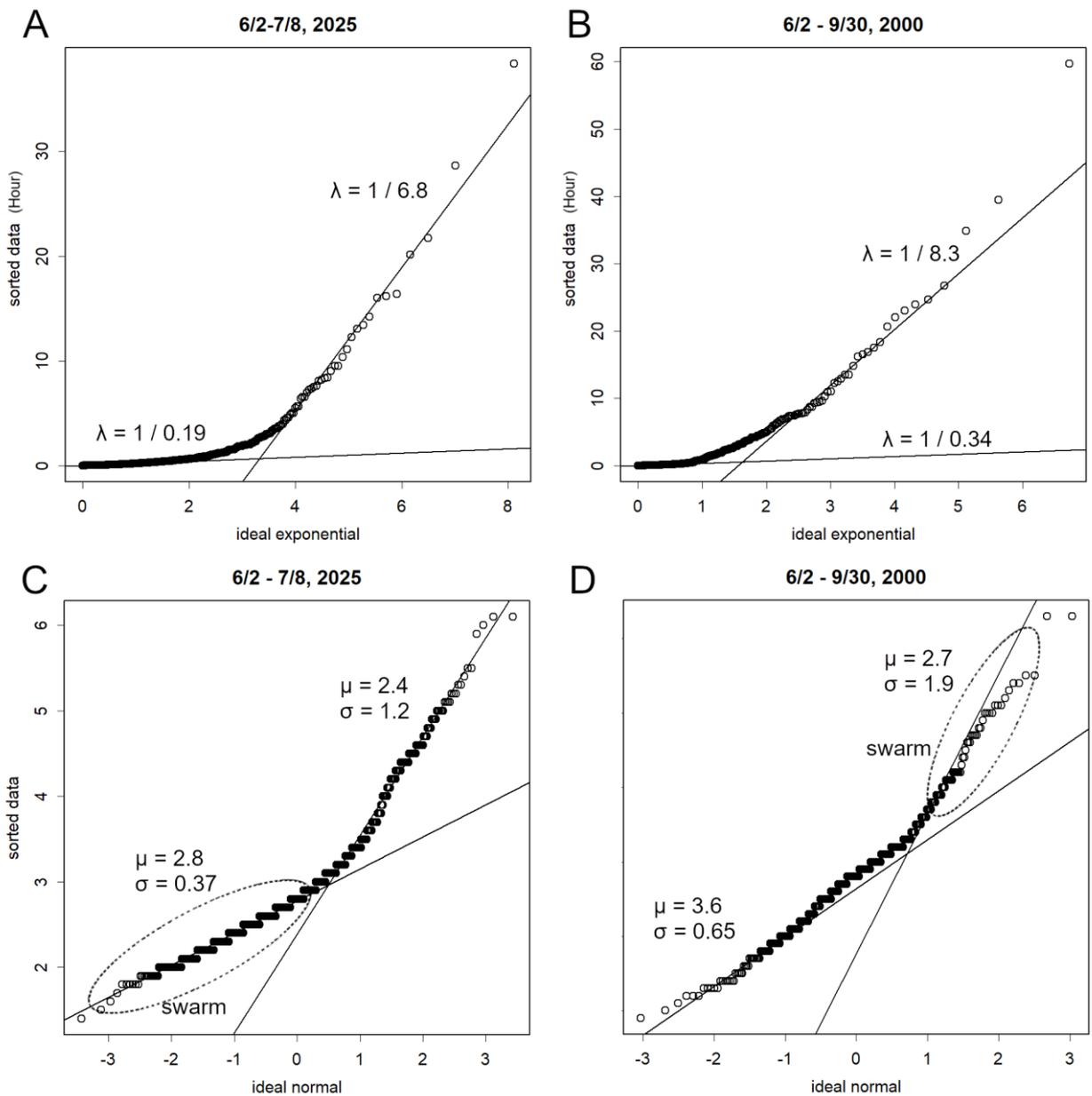

Figure 1. Distributional Analysis Using Quantile-Quantile (QQ) Plots

QQ plots comparing observed earthquake data (y-axis) with theoretical distributions (x-axis) for distributional validation. Each plot represents sorted data quantiles against ideal distribution quantiles.

**A**. Earthquake intervals for 2025 compared with ideal exponential distribution. Data encompass all Japanese seismic activity. Two distinct slopes are evident: high λ (λ = 1/0.19) corresponding to the Tokara Islands earthquake cluster and normal λ (λ = 1/6.8) representing background seismicity. The reciprocal of λ (1/λ) represents the mean earthquake interval for each population.

**B**. Earthquake intervals before and after the 2000 Miyake Island eruption. Pre-eruption activity follows exponential distribution ($\lambda = 1/8.3$), whilst post-eruption activity exhibits increased frequency ($\lambda = 1/0.34$), demonstrating characteristic swarm behaviour.

**C**. Magnitude distribution for 2025 compared with normal distribution. Two populations are distinguished: low scale ($\sigma = 0.37$, $\mu = 2.8$) associated with the Tokara earthquake swarm and higher variability ($\sigma = 1.2$, $\mu = 2.4$) representing regional background activity.

**D**. Magnitude distribution before and after the 2000 Miyake Island eruption compared with normal distribution. Pre-eruption magnitudes exhibit lower variability ($\sigma = 0.65$, $\mu = 3.6$), whilst post-eruption swarm activity demonstrates increased magnitude variability ($\sigma = 1.9$, $\mu = 2.7$), contrasting with the Tokara pattern.

On the morning of 8 July, the seismic swarm characteristics underwent subtle modifications (Figure 2A). The epicentral location shifted from the seabed between Akuseki Island and Takara Island to waters adjacent to Suwanose Island. From 9 July onwards, these parameters have exhibited a tendency to revert towards their pre-swarm values (Figure 2B and 2C), and the epicentral locations have returned to the original area.

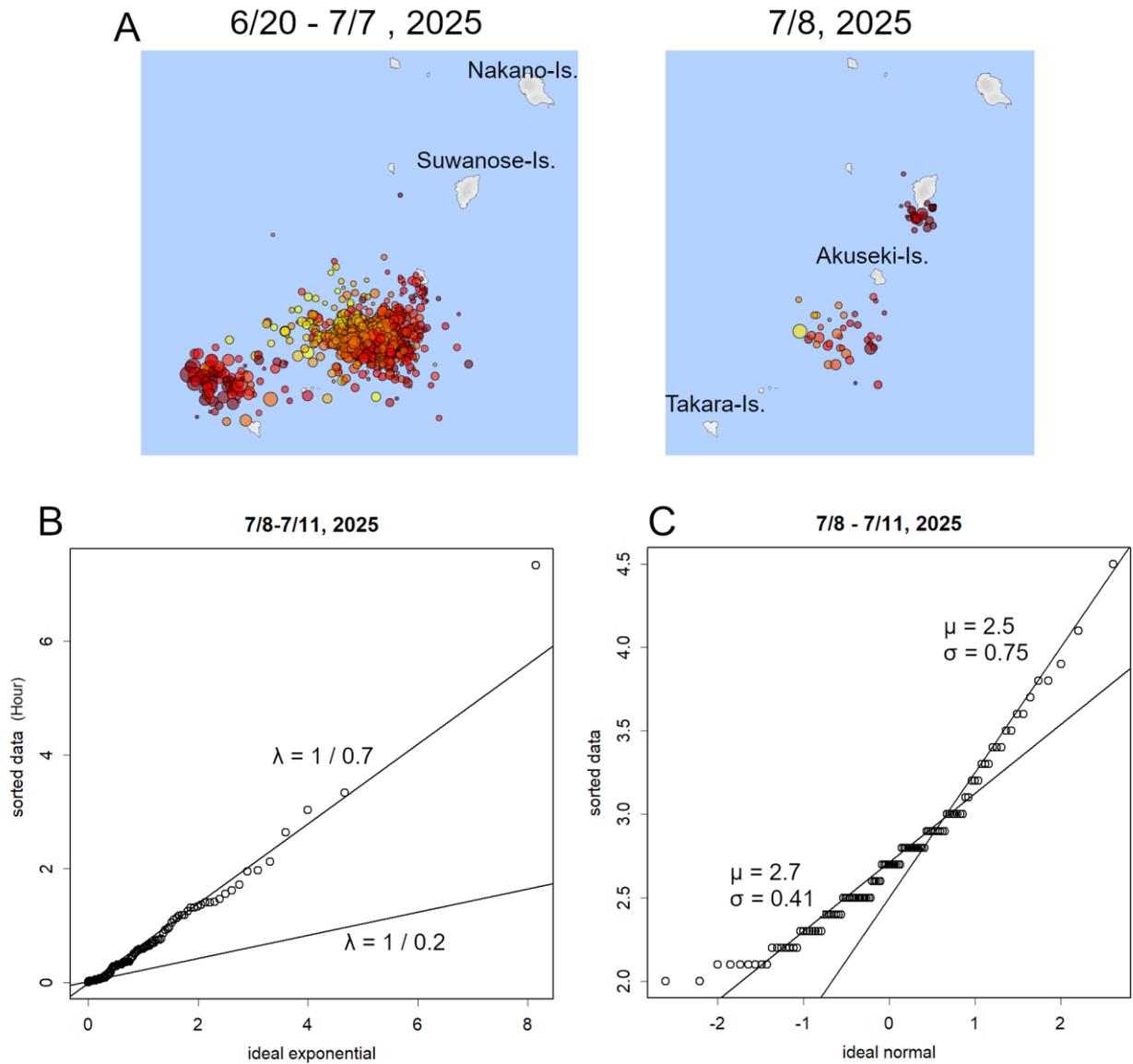

Figure 2. Temporal and Spatial Evolution of Seismic Activity

**A**. Epicentral migration during the earthquake swarm. Left panel shows epicentral distribution from 20 June to 7 July 2025, concentrated between Akuseki Island and Takara Island. Right panel displays epicentral locations on 8 July 2025, demonstrating northward migration towards Suwanose Island. The active volcano is situated at the centre of Suwanose Island, whilst Akuseki Island and Takara Island host geothermal springs but lack active volcanic centres.

**B**. Earthquake interval distribution (8-11 July 2025) compared with exponential distribution. Following epicentral migration, λ decreased (mean interval increased), showing two distinct populations: high-frequency events ($\lambda = 1/0.2$) and reduced-frequency events ($\lambda = 1/0.7$).

**C**. Magnitude distribution (8-11 July 2025) compared with normal distribution. Two populations are evident: the original low-variance population ($\sigma = 0.41$, $\mu = 2.7$) and an emerging higher-variance population ($\sigma = 0.75$, $\mu = 2.5$), suggesting transition of swarms.

The observed parameters for the current 2025 earthquake cluster demonstrate remarkable similarity to those recorded during the two cluster earthquakes of 2021 (Figure 3). When compared with theoretical ideal distributions, the observed distributions are statistically identical except in the right tail, where both intervals and magnitudes increase. The small right tail might indicate changes in the swarm location (Figure 2). Rather, it should be noted that the observed parameters in most of the data area demonstrate remarkable consistency between the current 2025 earthquake cluster and the two 2021 cluster events.

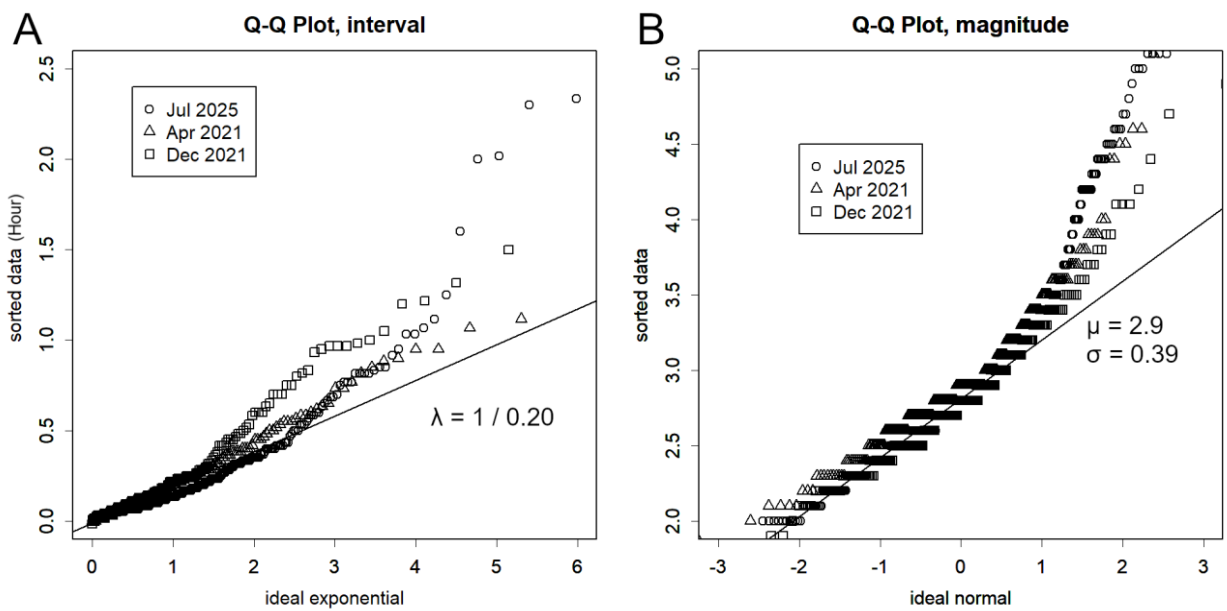

Figure 3. Comparative Analysis of Tokara Earthquake Swarms Across Multiple Years

QQ plots comparing earthquake swarm characteristics during three distinct periods: 29 June - 4 July 2025, 10-12 April 2021, and 4-7 December 2021.

**A**. Earthquake interval distributions compared with exponential distribution. The three swarm periods demonstrate remarkably similar statistical characteristics, with consistent exponential behaviour ($\lambda = 1/0.20$) during peak swarm activity, indicating identical underlying seismogenic processes.

**B**. Magnitude distributions compared with normal distribution. All three swarms exhibit nearly identical magnitude characteristics ($\mu = 2.9$, $\sigma = 0.39$), demonstrating consistent energy release patterns across different temporal episodes. The convergence of data points along the theoretical line confirms the reproducibility of swarm behaviour.

## DISCUSSION

In the month preceding the Tohoku earthquake, the magnitudes' $\sigma$ reached 1.3 whilst $\mu$ was significantly elevated at 3.6 (Figure S2, Appendix A). The frequency was exceptionally high, averaging 3.7 hours even when cluster earthquakes were excluded from the analysis. This trend was particularly pronounced in the more immediate pre-seismic period, creating conditions conducive to large-magnitude earthquakes (Konishi, 2025). The current situation in the Tokara Islands presents a markedly different pattern, characterised by lower $\sigma$ values (Figure 1). Consequently, there are presently no indications of an impending megathrust earthquake of Tohoku-class magnitude. Given that these data encompass the entirety of Japan, this suggests the absence of imminent major seismic activity across the Japanese archipelago.

The magnitude distribution of earthquakes occurring between Akuseki and Takara Islands exhibits consistently low $\sigma$ values, which have remained stable since 2021 (Figure 3B). Therefore, high-magnitude earthquakes are not anticipated in this region; although frequent low-magnitude events will continue and σ is expected to increase as the swarm activity diminishes, hence occasional moderate events may occur stochastically. The relatively low energy signatures observed may be attributed to magma movement, which requires less energy than plate displacement due to magma's lower viscosity compared to solid crustal material. In contrast, the earthquake swarms during the Miyake Island eruption may not have been primarily driven by magma movement but rather represented aftershock sequences from a major earthquake, with energy sources attributable to plate tectonics (Figure 1D).

The current earthquake swarm in the Tokara Islands is most likely caused by magma movement. The temporal correlation between volcanic activity at Suwanose Island and earthquake swarms at this location in both 2021 and 2025, coupled with nearly identical statistical parameters (Figure 3), suggests the presence of a persistent magma conduit system and a geological structure predisposed to seismic activity. The prolonged nature of current volcanic activity corresponds with the extended duration of the earthquake swarm. However, magma flow patterns appear to vary temporally, potentially triggering swarm activity at alternative locations with correspondingly different seismic parameters (Figure 2). Given the ongoing volcanic activity at Suwanose Island, magma movement is likely to persist, with earthquake activity continuing until volcanic processes cease.

Whether this activity will remain confined to the existing eruption at Suwanose Island or result in the formation of new eruptive centres (Japan Coast Guard, 2013) remains uncertain. This uncertainty stems from the absence of published quantitative data regarding eruption magnitude (Hayakawa, 1993). Access to quantitative measurements of the 2021 eruption and their relationship to seismic activity would enable meaningful comparison with current conditions. Observations suggest potential changes in island positions, indicating the possibility of large-scale geological modifications (JMA, 2025a). For predictive purposes, the development of standardised methods for quantifying eruptive activity would be highly beneficial.

CONCLUSION

The seismic activity discussed herein has persisted over an extended period and has occurred with notable frequency. Nevertheless, its parameters and epicentral location closely resemble those documented in 2021, with a particularly low magnitude scale ($\sigma$). A recurrent phenomenon associated with these events is the eruption of Suwanose Island, suggesting a strong link between the earthquakes and magmatic movement related to volcanic activity. Given the consistently small magnitude $\sigma$, the likelihood of a significant seismic event appears low. Furthermore, no precursory signals indicative of a large-scale earthquake, such as those observed prior to the 2011 Tohoku earthquake, have been detected.

APPENDICES

Appendix A: Supplementary figures.

REFERENCES

Gutenberg, B., and C. F. Richter, 1944, Frequency of Earthquakes in California: Bulletin of the Seismological Society of America, **34**, no. 4, 185-188.

Hayakawa, Y., 1993, Proposal for eruption magnitude: Kazan (Japanese), **38**, no. 6, 223-226. http://dx.doi.org/10.18940/kazan.38.6_223.

JAMSTEC, 2025, Kikai Caldera Comprehensive Survey, https://www.jamstec.go.jp/rimg/j/research/kikaicaldera/., accessed 7/10 2025.

JMA, 2025a, Assessment of seismic activity in the waters near the Tokara Islands, https://www.static.jishin.go.jp/resource/monthly/2025/20250703_tokara_2.pdf., accessed 7/10 2025.

JMA, 2025b, Assessment of seismic activity off the coast of the Tokara Islands, https://www.static.jishin.go.jp/resource/monthly/2025/20250703_tokara_2.pdf., accessed 7/10 2025.

JMA, 2025c, Nankai Trough Earthquake, https://www.jma.go.jp/jma/kishou/know/jishin/nteq/index.html., accessed 7/10 2025.

JMA, 2025d, Weather, earthquakes, volcanoes, ocean knowledge - Volcanoes, https://www.jma-net.go.jp/fukuoka/jikazan/kazan_hanasi.html., accessed 7/10 2025.

JMA, 2025e, Summary of seismic activity for each month, https://www.data.jma.go.jp/eqev/data/gaikyo/., accessed 7/10 2025.

JMA, 2025f, Epicenter location, https://www.jma.go.jp/bosai/map.html#10/29.428/129.559/&contents=hypo., accessed 7/10 2025.

Japan Coast Guard, 2013, The submarine volcano in TOKARA Islands Volcanic Eruption Prediction Liaison Committee Bulletin (JMA), **115**, 235-236. https://www.jma.go.jp/jma/kishou/shingikai/ccpve/Report/115/kaiho_115_37.pdf., accessed 7/10 2025.

Konishi, T., 2025, Seismic Pattern Changes Before the 2011 Tohoku Earthquake Revealed by exploratory data analysis: Interpretation, **13**, no. 4., in printing.

Nishimura, T., 2021, Volcanic eruptions are triggered in static dilatational strain fields generated by large earthquakes: Scientific Reports, **11**, no. 1, 17235. http://dx.doi.org/10.1038/s41598-021-96756-z.

R Core Team, 2025, R: A language and environment for statistical computing: R Foundation for Statistical Computing., accessed 7/10 2025.

Survey, B. G., 2025, How volcanoes form, https://www.bgs.ac.uk/discovering-geology/earth-hazards/volcanoes/how-volcanoes-form-2/., accessed 7/10 2025.

        Yamada, M., et al, 2020, Tsunami deposits from the Kikai Caldera eruption

found in the coastal lowlands of Tainohama, Tokushima Prefecture, and numerical simulations: Historical Earthquakes, **35**, 275.

https://www.histeq.jp/kaishi/HE35/HE35_275_Yamada.pdf.

## Appendix A: Supplementary Figures

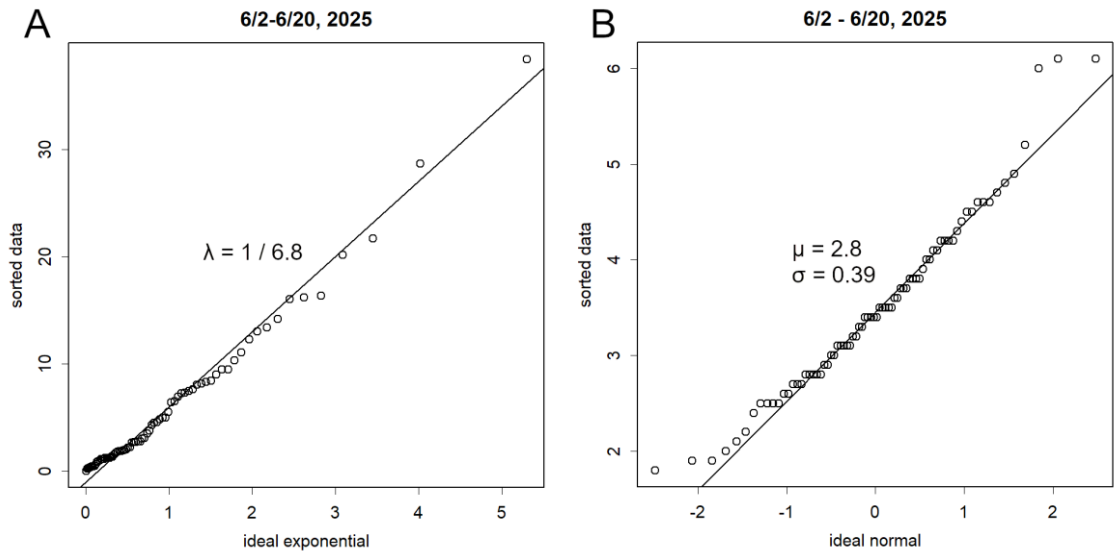

Figure S1. Statistical distributions before the earthquake swarm. The Q-Q plots show straight lines, indicating homogeneous distributions that follow theoretical models. A. Earthquake interval (exponential distribution, $\lambda = 1/6.8$). B. Magnitude (normal distribution, $\mu = 2.8$, $\sigma = 0.39$).

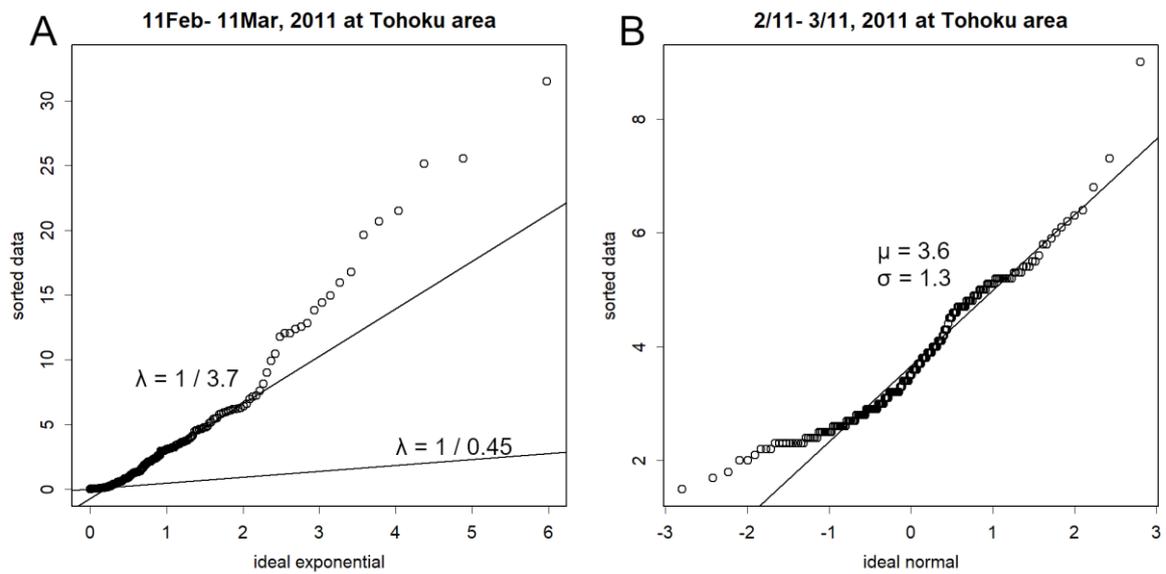

Figure S2. Statistical distributions during the month preceding the Tohoku

earthquake. A. Time interval. The parameter $\lambda$ is large, even when clustered earthquakes are excluded. B. Magnitude. Both $\mu$ and $\sigma$ show increased values.

This facilitated the repeated occurrence of larger magnitude earthquakes.